\documentclass[preprint,12pt]{elsarticle}
\usepackage{amsmath}
\usepackage{amssymb}
\usepackage[pdftex,colorlinks]{hyperref}
\usepackage{multirow}
\usepackage{enumerate}
\usepackage{cases}
\usepackage{extarrows}

\makeatletter

\newcommand {\Rmnum} [1] {\expandafter \@slowromancap \romannumeral #1@}
\makeatother

\newtheorem{Protocol}{Protocol}

\biboptions{compress}

\begin{document}

\begin{frontmatter}

\title{Counterfactual Quantum Bit Commitment}


%\author[label1]{Li Yang\corref{1}}\ead{yang@is.ac.cn}
%\cortext[1]{Corresponding author.}
%\ead{yang@is.ac.cn}
\author{Ya-Qi Song$^{1,2,3}$\address{1,2,3}, Li Yang$^{1,2,3}$\address{1,2,3}\corref{cor1}}
\cortext[cor1]{Corresponding author(emali:yangli@iie.ac.cn)}
\address{$^1$ State Key Laboratory of Information Security, Institute of Information Engineering,
Chinese Academy of Sciences, Beijing 100093, China\\
$^2$ Data Assurance and Communication Security Research Center,Chinese Academy of
Sciences, Beijing 100093, China\\
$^3$ School of Cyber Security, University of Chinese Academy of Sciences, Beijing 100049, China \\}

\begin{abstract}
We propose a framework of bit commitment protocol using a comparison scheme and present a compound comparison scheme based on counterfactual cryptography. Finally, we propose a counterfactual quantum bit commitment protocol. In security analysis, we give the proper security parameters for counterfactual quantum bit commitment and prove that intercept attack and intercept/resend attack are ineffective attack for our protocol. In addition, we explain that counterfactual quantum bit commitment protocol cannot be attacked with no-go theorem attack by current technology.
\end{abstract}


\end{frontmatter}
\section{Introduction}
The bit commitment (BC) scheme is a two-party protocol which  plays a crucial role in constructions of multi-party protocols. BC scheme includes two phases. In the commit phase, Alice commits to $b$ ( $b=0$ or $b=1$ ) and sends a piece of evidence to Bob. In the opening phase, Alice unveils the value of $b$ and Bob checks it with the evidence. A BC scheme has the following security properties. (i) Concealing. Bob cannot know the commitment bit $b$ before the opening phase. (ii) Binding. Alice cannot change the commitment bit after the commit phase. A BC scheme is unconditionally secure if and only if there is no computational assumption on attacker's ability and it satisfies the properties of concealing and binding.

The concept of BC was first proposed by Blum in \cite{Blum82}. With the development of quantum cryptography, the first quantum bit commitment (QBC) scheme was proposed in 1984 \cite{BB84} but unfortunately the binding security of the scheme can be attacked by entangled states. Then  a well-known QBC scheme was presented \cite{BCJL}, which is usually referred to as BCJL scheme and was once believed as a provably secure scheme. However, Mayers found that the BCJL scheme was insecure \cite{Mayers96}. Later, Mayers, Lo and Chau separately present no-go theorem and prove that the unconditional secure QBC protocol is impossible \cite{Mayers97,Lo97,BCMno-go97}.

However, the framework of the theorem may not cover all the types of QBC protocols.
Some QBC protocols against no-go theorem type attack have been proposed. Using special relativity, the relativistic QBC protocols are proposed by Kent \cite{Kent99,Kent05,Kent12}. Using the physical hypothesis, the bounded-quantum-storage model \cite{DF05,DD07}, and noisy-storage model \cite{WS08,NJ12,KW12} are presented.

In this paper, we first construct a universal framework for BC protocol. A comparison protocol is invoked in the framework. Then we propose the comparison protocol based on counterfactual quantum cryptography(N09)\cite{CQKD}. Finally, a counterfactual quantum bit commitment protocol (CQBC) is presented. In this CQBC protocol, Bob sends the states and Alice only receives some of the states. In the ideal protocol, Bob sends a single photon and he obviously knows whether Alice receives the photon. Alice's traditional attack based on no-go theorem needs to send or return the states to Bob. In this protocol, once she gets the states sent by Bob, Bob knows her choice and she cannot change the bit anymore. In addition, Alice's operation is to control the macroscopic device $SW$. It cannot be realized by quantum states, which is an important reason why Alice cannot perform no-go theorem type attack.
% It is a distinctive QBC protocol and we have had no idea of no-go theorem type attack so far. In this paper, we propose a special channel to give a framework of QBC protocol and present the counterfactual QBC protocol. Then we analyze concealing and binding security of the protocol.
\section{Preliminary}
\begin{figure}[!h]
\centering
\includegraphics[width=0.9\textwidth]{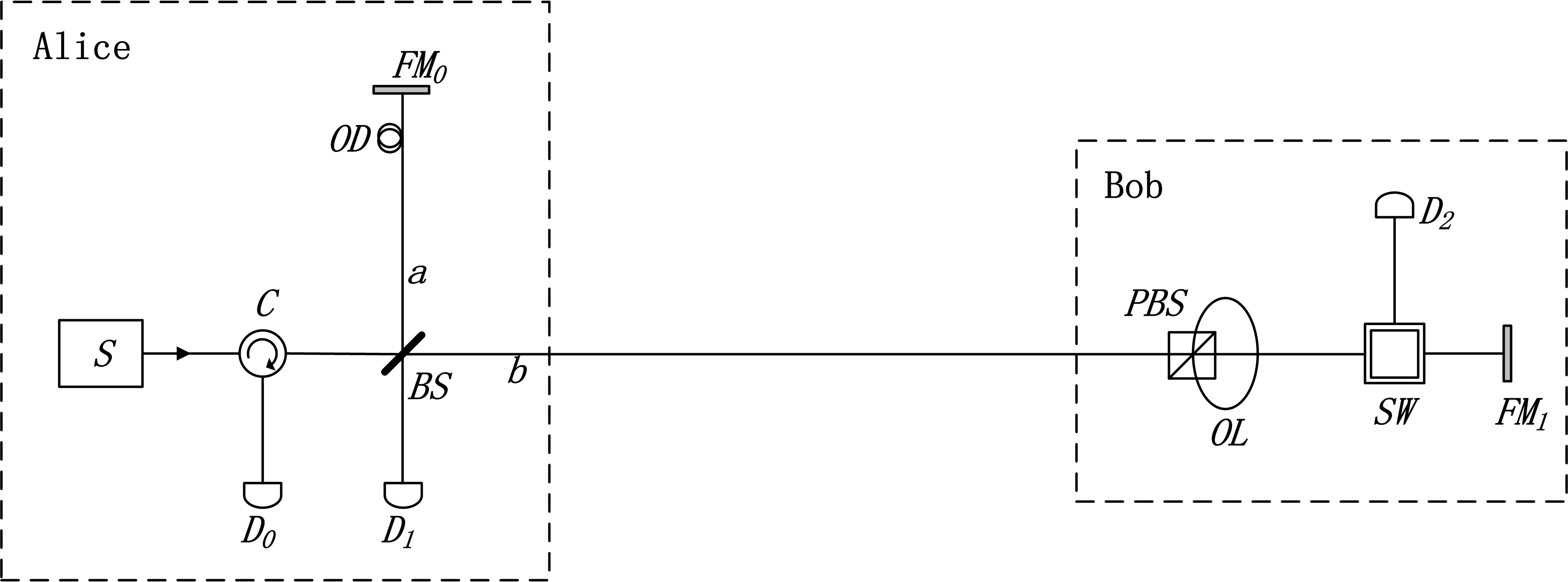}
\caption{The architecture of the N09 QKD protocol. The setup is a modification based on Michelson-type interferometer. The single photon source $S$ emits a optical pulse containing only one photon. Then the pulse is transmitted through the optical circulator $C$ and split into two pulses by the beam splitter $BS$. The two light paths $a$ and $b$ are the arms of the Michelson-type interferometer, and the length of the path $a$ is adjusted by an optical delay $OD$. The pulse transmitted through path $a$ is reflected by the Faraday mirror $FM_0$ and back to $BS$. The pulse transmitted through path $b$ travels to Bob's site. }
\label{fig:QKD}       % Give a unique label
\end{figure}

%\textbf{Counterfactual quantum cryptography.}
Noh proposed a special QKD protocol (N09)\cite{CQKD}, in which the particle carrying secret information is not transmitted through the quantum channel. Fig.~\ref{fig:QKD} shows the architecture of the QKD protocol.
In the QKD protocol, Alice randomly encodes horizontal-polarized state $|H\rangle$ as the bit value $"0"$ or vertical-polarized state $|V\rangle$ as the bit value $"1"$ and sends the state by the single photon source $S$. When Bob's bit value is the same as Alice's, the optical switch $SW$ controlled in the correct time. In this case, the interference is destroyed and there are three occasions for the single photon. Suppose the reflectivity and transmissivity of the $BS$ are $R$ and $T$, where $R+T=1$. The probabilities of detectors are as follows. (i) Detector $D_0$ clicks with the probability of $R^2$. The photon travels via path $a$ and then is reflected by the $BS$ again. (ii) Detector $D_1$ clicks with the probability of $RT$. The photon travels via path $a$ and then pass through the $BS$. (iii) Detector $D_2$ clicks with the probability of $T$. The photon travels via path $b$ and is controlled by the $SW$ to reach the detector $D_2$. When Bob's bit value is different from Alice's, the setup is a Michelson-type interferometer and the detector $D_0$ clicks. Alice and Bob only remain the bit in the event that the detector $D_1$ clicks alone to be the shared keys. The other events are used for eavesdropping detection.
The security of N09 protocol has been proved. In \cite{securityQKD10}, Yin et al. proposed an entanglement distillation protocol equivalent to the N09 protocol. Then give a strict security proof assuming that the perfect single photon source is applied and Trojan-horse attack can be detected. In 2012, Zhang et al. give a more intuitive security proof against the general intercept-resend attacks
\cite{ZW2012}.

\section{A Framework of Bit Commitment Protocol}

BC is a two-party cryptographic protocol. In the commit phase, one party Alice commits to the other party Bob to a bit $b$ by sending a piece of evidence. In the opening phase, Alice announces the value of $b$ and Bob verifies whether it is indeed the commitment bit. We give a framework to construct BC protocol. The BC scheme which satisfies this framework could be secure by selecting appropriate security parameters.
\begin{Protocol}
  \emph{\textbf{The framework of bit commitment protocol}}
~\\
Commit Phase:
 \begin{enumerate}
   \item Alice and Bob agree on two security parameters $m$ and $n$.
   \item Alice chooses a random bit $b \in{\{0,1\}}$ as her commitment bit. Then she generates $m$ random bit strings according to the value of $b$. Each sequence consists $n$ bits, which can be represented as $a^{(i)}\equiv ({a_1^{(i)}}{a_2^{(i)}}...{a_n^{(i)}}) \in{\{0,1\}}^{n}$, $i=1, 2,..., m$. Each sequence satisfies ${a_1^{(i)}}\oplus{a_2^{(i)}}\oplus...\oplus{a_n^{(i)}}=b$.
   \item Bob generates $m$ bit strings randomly and uniformly with the length of $n$. Each sequence is represented as $b^{(i)}\equiv ({b_1^{(i)}}{b_2^{(i)}}...{b_n^{(i)}}) \in{\{0,1\}}^{n}$.
   \item Alice and Bob invoke another particular protocol to give some evidence of commitment to Bob. In this step, Bob compares $b_j^{(i)}$ with $a_j^{(i)}$ bit-by-bit and knows $b_j^{(i)}=a_j^{(i)}$, $b_j^{(i)}\neq a_j^{(i)}$, or nothing. For each bit-comparison, Bob could confirm the value of Alice's bit with a probability $p$ and Alice knows that Bob confirms her bit with a probability $q$, where $0\leq q<p<1$.
 \end{enumerate}
~\\Opening Phase:
 \begin{enumerate}
   \item Alice reveals the bit $b$, the $m$ sequences $({a_1^{(i)}}{a_2^{(i)}}...{a_n^{(i)}})$, $i=1, 2,..., m$ to Bob.
   \item Bob verifies whether ${a_1^{(i)}}\oplus{a_2^{(i)}}\oplus...\oplus{a_n^{(i)}}=b$, and whether Alice's opening results consistent with the bits he knows. If the consistency holds, he admits Alice's commitment value as $b$.
 \end{enumerate}
\end{Protocol}

\section{Counterfactual Quantum Bit Commitment}
There is a particular protocol invoked in Step 4 of Protocol 1. In this section, we first construct the two-party protocol based on counterfactual cryptography. The aim of the two-party protocol is to realize the comparison bit by bit with a fix probability. Then invoke the comparison protocol to give the CQBC protocol.
\begin{figure}[h]
\centering
\includegraphics[width=0.9\textwidth]{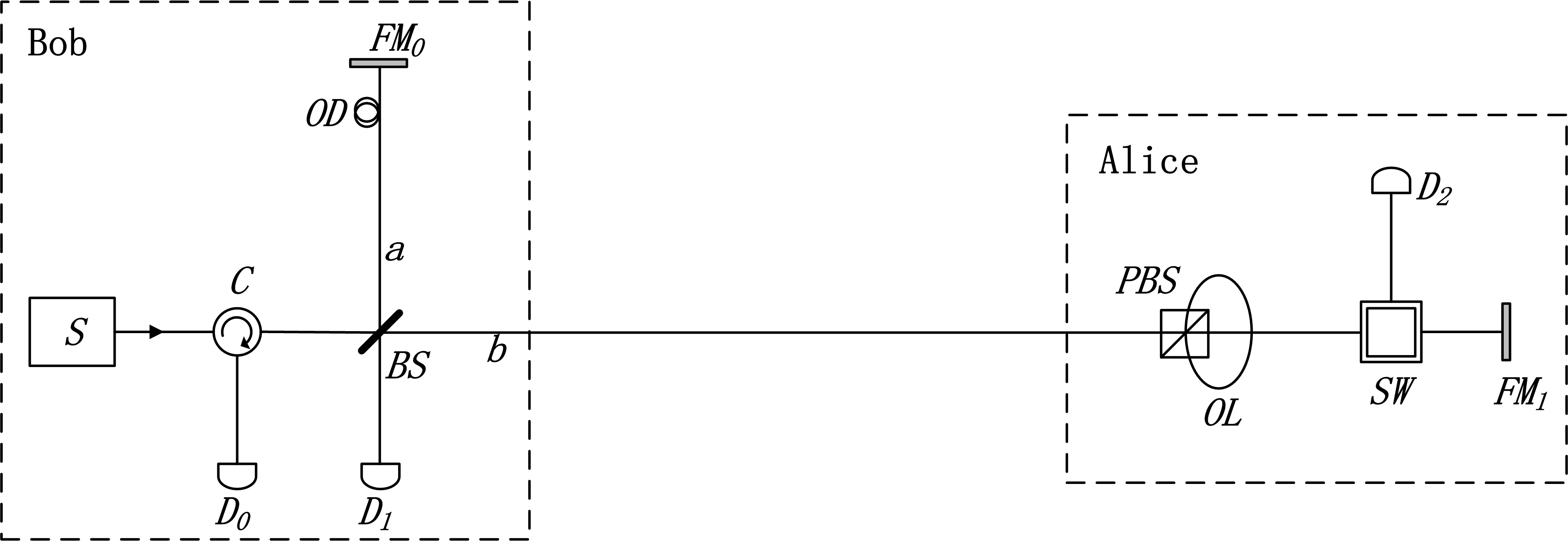}
\caption{The architecture of Protocol 2 and Protocol 3. The difference between this architecture with that of Fig.\ref{fig:QKD} is that Bob is the sender in this architecture. }
\label{fig:QBC}       % Give a unique label
\end{figure}
\begin{Protocol}
  \emph{\textbf{Comparison based on counterfactual cryptography}}
 \begin{enumerate}
   \item Alice and Bob set up devices according to Fig.~\ref{fig:QBC}, where the beam splitter $BS$ is a half transparent and half reflecting mirror.
   \item Alice and Bob perform a test to determine the time parameters. Bob sends a series of states $|H\rangle$ or $|V\rangle$ to Alice and tells her what the states are before sending. Then Alice tries to control the optical switch $SW$ in proper time to make detector $D_0$, $D_1$, $D_2$ click, respectively. Through this test, three time parameters could be determined. That are, $\Delta t_0$: the time that the states spend from the source $S$ through the polarizing beam splitter $PBS$ to the optical switch $SW$; $\Delta t_1$: the time that the states spend from the source $S$ through the optical loop $OL$ to the optical switch $SW$; $\Delta t_2$: the time that the states spend from the source $S$, reflected by $FM_1$ to Bob's site again.
   \item Alice and Bob decide on a series of time instants $t_1^{(i)}, t_2^{(i)},...,t_n^{(i)}$, where $i=1,2,...m$. Bob generates his comparison bits string $(b_1^{(i)}b_2^{(i)}...b_n^{(i)})\in \{0,1\}^n$ and sends the corresponding states $|\Psi_{b_j^{(i)}}\rangle$ at the time $t_j^{(i)}$, where $|\Psi_{0}\rangle=|H\rangle$ and $|\Psi_{1}\rangle=|V\rangle$.
   \item Alice generates her comparison bits string $(a_1^{(i)}a_2^{(i)}...a_n^{(i)})\in \{0,1\}^n$ and controls the optical switch $SW$ in the corresponding time. When $a_j^{(i)}=0$, she controls $SW$ at the time $t_j^{(i)}+\Delta t_0$; When $a_j^{(i)}=1$, she controls $SW$ at the time $t_j^{(i)}+\Delta t_1$.
   \item Alice and Bob record the response of the detector $D_2$, $D_0$, $D_1$ as $(\alpha_1^{(i)}\alpha_2^{(i)}$ $...\alpha_n^{(i)})\in \{0,1\}^n$, $(\beta_{01}^{(i)}\beta_{02}^{(i)}...\beta_{0n}^{(i)})\in \{0,1\}^n$, $(\beta_{11}^{(i)}\beta_{12}^{(i)}...\beta_{1n}^{(i)})\in \{0,1\}^n$, respectively. $\alpha_j^{(i)}, \beta_{0j}^{(i)}, \beta_{1j}^{(i)}=0$ denotes that there is no click in the related detector. $\alpha_j^{(i)}, \beta_{0j}^{(i)}, \beta_{1j}^{(i)}=1$ denotes the related detector clicks. Note that as long as the detectors do not click in the correct time, they record the result ``0''. For example, if Bob's detectors $D_0$ and $D_1$ have not clicked until $t_j^{(i)}+\Delta t_2$, he records $\beta_{0j}^{(i)}=\beta_{1j}^{(i)}=0$.
 \end{enumerate}
\end{Protocol}

\begin{Protocol}
\emph{\textbf{Counterfactual bit commitment}}
~\\
Commit Phase:
\begin{enumerate}
\item Alice and Bob set up devices according to Fig.~\ref{fig:QBC}, where the beam splitter $BS$ is a half transparent and half reflecting mirror. They share two security parameters $m$ and $n$.
\item Alice chooses a random bit $b \in{\{0,1\}}$ as her commitment bit. Then she generates $m$ random bit strings according to the value of $b$. Each sequence consists $n$ bits, which can be represented as $a^{(i)}\equiv ({a_1^{(i)}}{a_2^{(i)}}...{a_n^{(i)}})\in{\{0,1\}}^{n}$, $i=1, 2,..., m$. Each sequence satisfies ${a_1^{(i)}}\oplus{a_2^{(i)}}\oplus...\oplus{a_n^{(i)}}=b$.
\item Bob generates $m$ bit strings randomly and uniformly with the length of $n$. Each sequence is represented as $b^{(i)}\equiv ({b_1^{(i)}}{b_2^{(i)}}...{b_n^{(i)}})\in{\{0,1\}}^{n}$.
\item Alice and Bob decide on a series of time instants $t_1^{(i)}, t_2^{(i)}, ..., t_n^{(i)}$ and $\bigtriangleup t$, where $\bigtriangleup t$ is the time a photon transfers from the beam splitter $BS$ to the optical switch $SW$ through the polarizing beam splitter $PBS$ without the optical loop $OL$. Bob sends $|\Psi_{b_j^{(i)}}\rangle$ at the time $t_j^{(i)}$ while Alice controls the switch with bit $a_j^{(i)}$. $|\Psi_{0}\rangle=|H\rangle$ and $|\Psi_{1}\rangle=|V\rangle$ represent the horizontal-polarized state and the vertical-polarized state, respectively.
\item Alice and Bob record the time and response of their detectors. For each sequence of states, Alice verifies whether the detection of $D_2$ is around $n/4$. If the proportion is incongruent, abort the protocol.
\end{enumerate}
~\\Opening Phase:
\begin{enumerate}
\item Alice reveals the bit $b$, the $m$ sequences $({a_1^{(i)}}{a_2^{(i)}}...{a_n^{(i)}})$, $i=1, 2,..., m$ and the response of her three detectors to Bob.
\item Bob verifies whether ${a_1^{(i)}}\oplus{a_2^{(i)}}\oplus...\oplus{a_n^{(i)}}=b$, and the response of all the detectors agree with the state $|\Psi_{b_j^{(i)}}\rangle$. If the consistency holds, he admits Alice's commitment value as $b$.
\end{enumerate}
\end{Protocol}

\section{Security Analysis}
\subsection{Security of BC Model}
We present a framework to construct BC protocol in Protocol 1. For each bit-comparison, Bob confirms the value of Alice's bit with a probability $p$ and Alice knows that Bob confirms her bit with a probability $q$, where $0\leq q<p<1$. $p>0$ means that Bob has a piece of evidence. Since $p<1$, Bob cannot know all of Alice's bits correctly. By choosing appropriate security parameters $n$, the protocol can satisfy the concealing security.  If Alice tries to alter one bit in the opening phase, her best choice is to change the bit she cannot distinguish whether Bob knows with a probability of $1-q$. In fact, there are around $(1-p)n$ qubits Bob cannot judge. If $p=q$, Alice can accurately alters the bit in part that Bob really does not know without detection. If $q<p$, the range of bits that can be altered by Alice is larger than that  Bob cannot distinguish, and her attack may be caught. Therefore, the conditions $0\leq q<p<1$ is the necessary condition of the binding security.

\subsubsection{Binding of BC Model}
If Alice tries to attack the binding of the protocol, she has to alter odd bits for each sequence in the opening phase. In each sequence, she can distinguish that around $qn$ bits are confirmed by Bob. Alice's optimal strategy is to alter one bit in the range of the other $(1-q)n$ bits. Among the $(1-q)n$ bits, only $(1-p)n$ bits are not known by Bob. Therefore, the probability that Alice alters one bit without detection is
\begin{equation}
  p(Aalter)=\frac{(1-p)n}{(1-q)n}=\frac{1-p}{1-q}.
\end{equation}
Then in $m$ sequences, the probability of changing the commitment bit without detection is $p(Aatler)^m$. Since $p(Aalter)<1$, $p(Aatler)^m$ can be exponentially small and the protocol can satisfy the binding security by choosing appropriate security parameter $m$.

\subsubsection{Concealing of BC Model}
For each bit, Bob confirms the value with a probability $p$. In some particular conditions, Bob may have a larger probability $p'$ to guess the value correctly, which can be seen in Section 5.2. For a sequence of qubits, Bob makes sure the commitment value with a probability of $p'^n$. Given $m$ qubit strings, the probability that Bob has no idea about the commitment value is $(1-p'^n)^m$.  Define $\varepsilon$ as the probability that Bob ascertains the commitment value,
\begin{equation}
  \varepsilon\equiv1-\left(1-p'^n\right)^m.
\end{equation}
If Bob does not confirm the commitment value from the protocol, he just guess with a probability of $1/2$. Therefore, the probability that Bob obtains the right commitment value is
\begin{equation}
  p(Bknows)=\varepsilon+\frac{1-\varepsilon}{2}=\frac{1}{2}+\frac{\varepsilon}{2}.
\end{equation}
Then the advantage of Bob breaking the concealing security is
\begin{equation}\label{pright}
  \left|p(Bknows)-\frac{1}{2}\right|=\frac{\varepsilon}{2}=\frac{1}{2}-\frac{(1-p'^n)^m}{2}.
\end{equation}
Since $0<p'<1$, then
\begin{equation}
  \left|p(Bknows)-\frac{1}{2}\right|\simeq\frac{1}{2}-\frac{1-mp'^n}{2}=mp'^n.
\end{equation}
$\left|p(Bknows)-\frac{1}{2}\right|$ can be exponentially small and the protocol can satisfy the concealing security by choosing appropriate security parameters $m$ and $n$.

\subsection{Analysis of Comparison Protocol}
Bob sends single-photon states $|H\rangle$ and $|V\rangle$ representing the bit value ``0'' and ``1''. The initial states after the beam splitter $BS$ become
\begin{equation}
  \begin{aligned}
    &|\phi_0\rangle=\sqrt{t}|0\rangle_a|H\rangle_b+i\sqrt{r}|H\rangle_a|0\rangle_b,\\
    &|\phi_1\rangle=\sqrt{t}|0\rangle_a|V\rangle_b+i\sqrt{r}|V\rangle_a|0\rangle_b,
  \end{aligned}
\end{equation}
where $a$ and $b$ represent the path towards Bob's Faraday mirror $FM_0$ and the path towards Bob's site, respectively. $t$ and $r$ are the transmissivity and the reflectivity of the $BS$. Both $|\phi_0\rangle$ and $|\phi_0\rangle$ can be denoted as Fock state $|\phi\rangle=\sqrt{t}|0\rangle_a|1\rangle_b+i\sqrt{r}|1\rangle_a|0\rangle_b$.

When $a_j^{(i)}=b_j^{(i)}$, the state $|\phi\rangle$ collapses to one of the two states, $|0\rangle_a|1\rangle_b$ or $|1\rangle_a|0\rangle_b$ due to Alice's measurement with probability $t$ and $r$, respectively. The state $|1\rangle_a|0\rangle_b$ goes past the $BS$ again and becomes $\sqrt{t}|0\rangle_{0}|1\rangle_1+i\sqrt{r}|1\rangle_0|0\rangle_1$, where the subscript $0$ and $1$ represent the path containing $D_0$ and $D_1$, respectively. Therefore, the total probability that $D_0$ detects the photon is $r^2$ and the probability that $D_1$ detects the photon is $rt$.

When $a_j^{(i)}\neq b_j^{(i)}$, one of the path introduces $\pi$ phase and the initial state becomes $\sqrt{t}|0\rangle_a|1\rangle_b-i\sqrt{r}|1\rangle_a|0\rangle_b$. Then the state passes the $BS$ again and becomes
\begin{equation}
  \begin{aligned}
    &\sqrt{t}|0\rangle_a|1\rangle_b-i\sqrt{r}|1\rangle_a|0\rangle_b\\
    \xrightarrow{BS}&\sqrt{t}(\sqrt{t}|1\rangle_0|0\rangle_1+i\sqrt{r}|0\rangle_0|1\rangle_1)- i\sqrt{r}(\sqrt{t}|0\rangle_{0}|1\rangle_1+i\sqrt{r}|1\rangle_0|0\rangle_1)\\
    \xlongequal{\quad}&t|1\rangle_0|0\rangle_1+i\sqrt{rt}|0\rangle_0|1\rangle_1- i\sqrt{rt}|0\rangle_{0}|1\rangle_1+r|1\rangle_0|0\rangle_1\\
    \xlongequal{\quad}&|1\rangle_0|0\rangle_1.
  \end{aligned}
\end{equation}
It can be seen that when $a_j^{(i)}\neq b_j^{(i)}$, the photon is detected by $D_0$ with a probability $100\%$.

 \begin{table}[!h]
\caption{The detection probability of each detector.  $r$ and $t$ are the reflectivity and transmissivity of the beam splitter $BS$.}
\label{tab:1}       % Give a unique label
\centering
\begin{tabular}[!h]{|c|c|c|}
\hline\noalign{\smallskip}
~~ & $a_j^{(i)}\neq b_j^{(i)}$ & $a_j^{(i)}=b_j^{(i)}$ \\
\noalign{\smallskip}\hline\noalign{\smallskip}
$\beta_{0j}^{(i)}=1$     & $1$     & $r^2$ \\\hline
$\beta_{1j}^{(i)}=1$     & $0$ & $rt$ \\\hline
$\alpha_{j}^{(i)}=1$     & $0$                & $t$ \\
\hline
\end{tabular}
\end{table}
The detection probability of each detector are listed in Table \ref{tab:1}. When the detector $D_1$ or $D_2$ clicks (detector $D_0$ does not click), Bob confirms Alice's bit is the same as his. It can be seen that
\begin{equation}
  p=p(a_j^{(i)}= b_j^{(i)},\beta_{1j}^{(i)}=1)+p(a_j^{(i)}= b_j^{(i)},\alpha_{j}^{(i)}=1)
  =\frac{1}{2}(rt+t).
\end{equation}
When $D_0$ clicks, it can be seen that $p(a_j^{(i)}\neq b_j^{(i)})>p(a_j^{(i)}= b_j^{(i)})$. Although Bob cannot confirm the value of $a_j^{(i)}$, he can guess $a_j^{(i)}\neq b_j^{(i)}$ with a correct probability of $p(a_j^{(i)}\neq b_j^{(i)}|\beta_{0j}^{(i)}=1)$, where
\begin{equation}
 p(a_j^{(i)}\neq b_j^{(i)}|\beta_{0j}^{(i)}=1)=\frac{1}{1+r^2}.
\end{equation}
Then the probability that Bob guesses Alice's bit $a_j^{(i)}$ correctly is
\begin{equation}
  p'=p+p(a_j^{(i)}\neq b_j^{(i)},\beta_{0j}^{(i)}=1)=7/8.
\end{equation}
When the detector $D_2$ clicks, Alice confirms Bob has obtained her bit. Therefore,
\begin{equation}
  q=\frac{1}{2}t=1/4.
\end{equation}

\subsection{Security of Counterfactual Quantum Bit Commitment}
We have proved the security of the BC framework for fixed parameters $p$, $p'$ and $q$. Then for the comparison protocol based on counterfactual cryptography, we analyze the related parameters. In this section, we will analyze the possible attacks for the complex protocol. The schematic of Protocol 2 and Protocol 3 is simple that there exist only a few attacks. For Bob, he may attack by change the beam splitter with different parameters or send illegal states. For Alice, she has two kinds of attacks, i.e. intercept attack and intercept/resend attack. In addition, we discuss the reason why Alice can hardly apply the attack using no-go theorem.

\subsubsection{Bob's Cheating}
The emission device is in Bob's site. The general attacks are to send illegal states and change the device.

If Bob sends illegal single-photon states with different polarizations, such as $|+\rangle$ or $|-\rangle$, it just influences the photons transmitted or reflected by $PBS$. And it can never increase the probability $p$, which is an ineffective attack.

Bob may attack by sending illegal multi-photon states. When multiple photons are transferred  in the scheme, the number of photons detected by $D_2$ is larger than $n/4$. In Step 5 of Protocol 3, Alice verifies the detection of $D_2$ and this attack can be found by the check.

Bob may not using a standard half transparent and half reflecting mirror in the protocol. Assume the transmissivity of the illegal $BS$ is $t'$, then clicks of $D_2$ is around $t'/2$. Different $BS$ leads different clicks of $D_2$. This attack can also be detected by the check in Step 5 of Protocol 3.

\subsubsection{Alice's Cheating}
\textbf{Intercept attack.} When Alice performs intercept attack, the probability $q$ would be increased and she may have a larger probability of altering the commitment without detection. Then we will analyze whether it is an effective attack. Alice can control the optical switch $SW$ both at the time $t_j^{(i)}+\Delta t_0$ and $t_j^{(i)}+\Delta t_1$ to increase the probability $q$. However, if she intercepts all of the photons transmitted through the beam splitter $BS$, the number of photons detected by $D_2$ is around $n/2$ in a n-bit sequence. The obliviously wrong ratio can be detected by Bob. Therefore, Alice should only select a few of photons to intercept.

Assume Alice selects $n_0$ photons to intercept. She intercepts the photons both in the cases $a_j^{(i)}\neq b_j^{(i)}$ and $a_j^{(i)}= b_j^{(i)}$. When $a_j^{(i)}\neq b_j^{(i)}$, the number of photons detected by $D_2$ is $n_0$; the number of photons detected by $D_0$ is $n-n_0$. When $a_j^{(i)}= b_j^{(i)}$, the number of photons detected by $D_2$ is $n_0+(tn-n_0)=tn$; the number of photons detected by $D_1$ is $rtn$; the number of photons detected by $D_0$ is $r^2n$. Therefore, the total clicks for detectors $D_0$, $D_1$ and $D_2$ are
\begin{equation}
 \begin{aligned}
  &N(\beta_{0j}^{(i)}=1)=\frac{1}{2}(n-n_0)+\frac{1}{2}r^2n=\frac{5}{8}n-\frac{1}{2}n_0;\\
  &N(\beta_{1j}^{(i)}=1)=\frac{1}{2}rtn=\frac{1}{8}n;\\
  &N(\alpha_{j}^{(i)}=1)=\frac{1}{2}n_0+\frac{1}{2}tn=\frac{1}{4}n+\frac{1}{2}n_0.
 \end{aligned}
\end{equation}
The total clicks are $N(\beta_{0j}^{(i)}=1)+N(\beta_{1j}^{(i)}=1)+N(\alpha_{j}^{(i)}=1)=n$. When $\alpha_{j}^{(i)}=1$, Alice knows that Bob confirms her bit. Her optimal strategy is to alter one bit in the range of $n-N(\alpha_{j}^{(i)}=1)$ bits. Among $n-N(\alpha_{j}^{(i)}=1)$ bits, only $N(\beta_{0j}^{(i)}=1)$ bits are not confirmed by Bob. Therefore, the probability that Alice alters one bit without detection by this attack is
\begin{equation}
  p'(Aalter)=\frac{N(\beta_{0j}^{(i)}=1)}{n-N(\alpha_{j}^{(i)}=1)}=\frac{5n-4n_0}{6n-4n_0}.
\end{equation}
When Alice does not intercept, the probability of altering one bit without detection is $p(Aalter)=5/6$. It can be seen that $p'(Aalter)<p(Aalter)$. The intercept attack makes Alice detected by Bob with larger probability and it is not an effective attack.

\textbf{Intercept/resend attack.} When Alice performs intercept attack, the numerator and denominator of $p(Aalter)$ are both increased. Then it makes Alice detected by Bob with larger probability and it is not an effective attack. We will analyze another similar attack, i.e. intercept/resend attack.  Alice controls the optical switch $SW$ both at the time $t_j^{(i)}+\Delta t_0$ and $t_j^{(i)}+\Delta t_1$. When she detects each photon, she immediately sends another photon with the same polarization back to Bob's site. If Alice intercepts and resends all of the photons transmitted through the beam splitter $BS$, the numbers of the photons detected by $D_0$ and $D_1$ are the same, which is different from the original ratio and detected by Bob. Therefore, Alice should select only a few photons and resend them back.

Assume Alice selects $n'_0$ photons to intercept and resend. She intercepts and resends the photons both in the cases $a_j^{(i)}\neq b_j^{(i)}$ and $a_j^{(i)}= b_j^{(i)}$. When $a_j^{(i)}\neq b_j^{(i)}$, the number of photons detected by $D_2$ is $n'_0$; the number of photons detected by $D_1$ is $rn'_0$; the number of photons detected by $D_0$ is $n-n'_0+tn_0'$. When $a_j^{(i)}= b_j^{(i)}$, the number of photons detected by $D_2$ is $tn$; the number of photons detected by $D_1$ is $rtn+rn'_0$; the number of photons detected by $D_0$ is $r^2n+tn'_0$. Therefore, the total clicks for detectors $D_0$, $D_1$ and $D_2$ are
\begin{equation}
 \begin{aligned}
  &N'(\beta_{0j}^{(i)}=1)=\frac{1}{2}(n-n'_0+tn'_0)+\frac{1}{2}(r^2n+tn'_0)=\frac{5}{8}n;\\
  &N'(\beta_{1j}^{(i)}=1)=\frac{1}{2}rn'_0+\frac{1}{2}(rtn+rn'_0)=\frac{1}{8}n+\frac{1}{2}n'_0;\\
  &N'(\alpha_{j}^{(i)}=1)=\frac{1}{2}n'_0+\frac{1}{2}tn=\frac{1}{4}n+\frac{1}{2}n'_0.
 \end{aligned}
\end{equation}
Since Alice resends $n'_0$ photons, the total clicks are $N'(\beta_{0j}^{(i)}=1)+N'(\beta_{1j}^{(i)}=1)+N'(\alpha_{j}^{(i)}=1)=n+n'_0$.
Among $N'(\alpha_{j}^{(i)}=1)$ bits, there are $n'_0$ bits intercepted and resent by Alice. The indexes of intercepted bits are the same as that of resent bits. For these $n'_0$ bits, although Alice knows the related value of $b_j^{(i)}$, she has no idea whether the resent bit is detected by $D_0$ or $D_1$. Therefore, Alice do not know whether Bob confirms these $n'_0$ bits. When she changes her commitment, the altering range is $n-[N'(\alpha_{j}^{(i)}=1)-n'_0]$. Only $N'(\beta_{0j}^{(i)}=1)$ bits are not confirmed by Bob and Alice changes these bits would not be detected. Therefore, the probability that Alice alters one bit without detection by this attack is
\begin{equation}
  p''(Aalter)=\frac{N'(\beta_{0j}^{(i)}=1)}{n-[N'(\alpha_{j}^{(i)}=1)-n'_0]}=\frac{5n}{6n+4n_0}.
\end{equation}
It can be seen that $p''(Aalter)<p(Aalter)$. The intercept/resend attack makes Alice detected by Bob with larger probability and it is not an effective attack either.

\textbf{No-go theorem attack.}The frame of  no-go theorem is described as follows.
When Alice commits $b$, she prepares
\begin{equation}\label{no-go state}
|b\rangle=\sum_{i} \alpha_i^{(b)}\big|e_i^{(b)}\big\rangle_A\otimes \big|\phi_{i}^{(b)}\big\rangle_B,
\end{equation}
where $\big\langle e_i^{(b)}\big|e_j^{(b)}\big\rangle_A=\delta_{ij}$ while $\big|\phi_{i}^{(b)}\big\rangle_B$'s are not necessarily orthogonal to each other.
She sends the second register to Bob as a piece of evidence.
To ensure the concealing of the QBC protocol, the density matrices describing the second register are approximative. i.e.,
\begin{equation}
\label{concealing}
  Tr_A|0\rangle\langle0|\equiv\rho_0^B\simeq \rho_1^B\equiv Tr_A|1\rangle\langle1|.
\end{equation}
When Eq.~(\ref{concealing}) is satisfied, Alice can apply a local unitary transformation to rotate $|0\rangle$ to $|1\rangle$ without detection.

In Protocol 3, the quantum states are prepared by Bob and Alice has no original states. If Alice wants to attack using no-go theorem, she tries to perform a controlled unitary transformation instead of the protocol operation, which is inspired by \cite{Yang13}. The control bit in the transformation is entangled with the other register. That is, when Alice commits ``0'', the whole state is
\begin{equation}\label{state0}
 \begin{aligned}
   |0\rangle&=\frac{1}{2^{n-1}}\sum_{a_1^{(i)}\oplus...\oplus a_n^{(i)}=0}|a_1^{(i)}...a_n^{(i)}\rangle_AU_B(a_1^{(i)}...a_n^{(i)}) \bigotimes_{j=1}^{n}|\Psi_{b_j^{(i)}}\rangle_B\\
   &=\frac{1}{2^{n-1}}\sum_{a_1^{(i)}\oplus...\oplus a_n^{(i)}=0}|a_1^{(i)}...a_n^{(i)}\rangle_A [U_B(a_1^{(i)})|\Psi_{b_1^{(i)}}\rangle_B]\otimes...\otimes [U_B(a_n^{(i)})|\Psi_{b_n^{(i)}}\rangle_B]\\
   &=\frac{1}{2^{n-1}}\sum_{a_1^{(i)}\oplus...\oplus a_n^{(i)}=0}|a_1^{(i)}...a_n^{(i)}\rangle_A  |\Psi'_{b_1^{(i)}}\rangle_B\otimes...\otimes|\Psi'_{b_n^{(i)}}\rangle_B.
  \end{aligned}
\end{equation}
Similarly, when Alice commits ``1'', the whole state is
\begin{equation}\label{state1}
   |1\rangle
   =\frac{1}{2^{n-1}}\sum_{a_1^{(i)}\oplus...\oplus a_n^{(i)}=1}|a_1^{(i)}...a_n^{(i)}\rangle_A  |\Psi'_{b_1^{(i)}}\rangle_B\otimes...\otimes|\Psi'_{b_n^{(i)}}\rangle_B.
\end{equation}
Since the concealing of Protocol 3 can be satisfied, Alice can perform a local unitary transformation to rotate $|0\rangle$ to $|1\rangle$. However, two characters limit this attack can hardly work with current technology.
\begin{enumerate}
 \item In Protocol 3, the operation of Alice is to control the optical switch $SW$ at different time according to $a_j^{(i)}$. The $SW$ is a macrocosmic device. If Alice does not replace the macrocosmic optical switch, her attack operation is equivalent to exponential Schrodinger's cat, which is to use superposed quantum states to control the macrocosmic devices coherently. Since Schrodinger's cat has not been implemented yet, this kind of attack cannot be realized now.
 \item Through the above reason, the only way of performing no-go theorem attack is to replace the macrocosmic optical switch with microcosmic device. However, how to use microcosmic device to realize the function of $SW$ is unsolved and it is a question for the future research.

 %\item Bob sends single-photon in the protocol. Bob can judge whether Alice receives the states according to his detectors. Alice's cheating needs to receive the state sent by Bob. However, if she receives the state by detector $D_2$, it means that $a_j^{(i)}=b_j^{(i)}$ and Bob knows her bit. Any change on this bit will be detected by Bob.
\end{enumerate}

\subsection{Security Parameters}
In Section 5.3, we have analyze that Alice's intercept attack and intercept/resend attack cannot work. The probability that Alice alters one bit without detection is
\begin{equation}
  P(Aalter)=\frac{1-p}{1-q}=5/6.
\end{equation}
Then in the QBC scheme, the probability of changing the commitment bit without detection is $P(Aatler)^m$. When $m=70$, the probability that Alice breaks the binding security is approximate to $2.8\times 10^{-6}$.

In Step 5 of Protocol 3, Alice verifies the detection of $D_2$ and this check makes Bob cannot send illegal states or use illegal devices. The probability that Bob guesses Alice's bit $a_j^{(i)}$ correctly is limited to $p'=7/8$. And the advantage of Bob breaking the concealing security is
\begin{equation}\label{pright}
 \left|P(Bknows)-\frac{1}{2}\right|=\frac{1}{2}-\frac{(1-P_B^n)^m}{2}
\end{equation}
When $m=70$, $n=130$, the probability that Bob breaks the concealing security is approximate to $1.0\times 10^{-6}$.

To limit cheating probability around $10^{-6}$, $m=70$, $n=130$, is one pair of proper parameters. The values of security parameters can be set up according to different security requirement.

\section{Discussion}
There are two critical parameters $p$ and $q$ in Protocol 1, where $p$ is the probability that Bob confirm the value of Alice's bit and $q$ is the probability that Alice knows Bob confirms. Does that means the protocol is superluminal? Absolutely not! It can be seen in Protocol 2 and Protocol 3 the single photon is transferred to Alice's site and then the photon or no photon returns to Bob's site. Bob obtains the information according to the response of his detectors. There is an interactive process in the protocol. Actually, the interactive process, including quantum states interaction and classical information interaction, is necessary for the BC framework.

\section{Conclusion}
We first construct a universal framework for BC protocol using comparison scheme. Then we propose the comparison protocol based on counterfactual quantum cryptography. Finally, a CQBC protocol is presented. Then we analyze the security of three protocols and give the proper security parameters for CQBC protocol. For concealing security, we prove that cheating Bob sending illegal states and using illegal devices can be detected by Alice. For binding security, we prove that Alice's intercept attack and intercept/resend attack are both ineffective attack. No-go theorem attack can hardly be performed with current technology for two reasons: (i) If Alice uses the macroscopical optical switch, her attack
operation is equivalent to using superposed quantum states to control the macroscopical devices, which cannot be realized now; (ii) The way of using  microcosmic device to realize the function of $SW$ is an unsolved question to be researched in the future.

\section*{Acknowledgements}
This work was supported by National Science Foundation of China (Grant No.61672517).

\end{document}